\documentclass{article}

\input{tcilatex}

\begin{document}

\noindent \textbf{Using a Kinematic Definition of the Hubble Parameter to
Determine}

\textbf{\ }\ \ \ \ \ \ \textbf{the} \textbf{Cosmological Constant}$\ \mathbf{%
\Lambda =0}$ \textbf{in a Balanced Universe}

\bigskip

\ \ \ \ \ \ \ \ \ \ \ \ \ \ \ \ \ \ \ \ \ \ \ \ \ \ \ \ \ \ \ \ \ \ \ \ \ \
\ \ \ \ \ \ \ \ D. Savickas\footnote{%
Electronic address: david.savickas@wne.edu}

\ \ \ \ \ \ \ \ \ \ \ \ \ \ \ Department of Physics,\ Western New England \
University

\ \ \ \ \ \ \ \ \ \ \ \ \ \ \ \ \ \ \ \ \ \ \ \ \ \ \ \ \ \ \ Springfield,
Massachusetts 01119

\bigskip

The Hubble parameter is kinematically defined in terms of the positions and
velocities of all particles in a universe which may or may not be finite.
This definition is set equal to the Hubble parameter as defined in the
Friedman-Lema\^{\i}tre solution of general relativity, and which occurs
after the inflationary expansion has ended in the Guth model. Because a
coordinate system at rest relative to its local Hubble drift is a system in
which the cosmic background radiation is observed to be isotropic, it is
also an inertial system. Just before the first mass particles are created
within a pure radiation universe, there are no mass particles that exist
which can define $H$ or the inertial systems associated with the Hubble
drift. It will be shown that only a cosmological constant with a magnitude
of zero will allow radiation to form mass particles that have a total energy
which is independent of inertial systems and is equal to the equivalent
energy of their rest mass. Additional mass particles are continuously formed
from the radiation throughout the expanding universe after the initial
particles are created.

\bigskip

\textbf{I. Introduction}

The Hubble parameter $H$ and the recessional motion in an expanding universe
describes the receding velocity $v$ between any two galaxies and the
distance between them $r$ to be related to each other by the relation $v=Hr$%
. This occurs exactly when neither galaxy has a peculiar velocity which
would \ have been caused by a local force such as gravitation. Current
observations indicate that this expansion of the universe appears to be
either balanced with the velocity of the receding galaxies either completely
vanishing or approaching a very small magnitude as the distance between all
the galaxies approach infinity. This is a peculiar occurrence because it
would indicate that the initial recessional velocities of the galaxies would
be required to have a magnitude that provides a kinetic energy that is
exactly enough to allow it to overcome gravity and expand while coming to
rest at infinity with extremely little or no residual velocity. No physical
effect is known that could cause such a result.

The purpose of this paper will be to show that when inertial systems are
defined to be systems\ which are either at rest or moving with a constant
velocity relative to their local Hubble drift, they are kinematically
defined by the Hubble parameter. The standard relativistic expansion
occurring after the inflationary expansion in the Guth model [1] requires
that $\Lambda =0$ so that the expansion is exactly balanced. It will be
shown that this is required to occur when the Hubble parameter itself can be
kinematically defined as an explicit function of the positions and
velocities of all mass particles in the universe.

The Hubble parameter can only be defined in terms of mass particles and
cannot be defined in terms of photons for the following reason: Although
radiation density in general relativity causes gravitational decelerations
and can change the magnitude of $H$, it is only mass particles whose
positions and velocities alone can be used in a kinematic definition of the
Hubble parameter and not the positions or velocities of the radiation's
photons because they have a fundamentally different behavior. A mass
particle's velocity refers to the motion of its rest mass, and its motion
relative to other mass particles can\ be measured by the change in
interparticle distances between masses or relative to an observer, and does
not require reference to inertial systems. The velocity $c$ of a photon has
physical meaning only in reference to the local inertial system in which it
is moving, and observers at rest in different inertial systems which are
moving relative to each other see different frequencies and energies for a
photon which could be used to define the equivalent rest mass of the photon.
The definition of a distant photon's velocity $v$ cannot be defined in terms
of the rate of change of interparticle distances alone, but only in terms of
its peculiar velocity $c$ relative to the receding inertial system in which
it is located and which is then added to the recessional Hubble velocity $Hr$
of the system itself so that an approaching photon's velocity is $v=Hr-c$.
The photon's velocity depends on $H$ and therefore cannot be used in turn
within the definition of the magnitude of $H$ which determines the inertial
system's receding velocity. It is for this reason that the kinematic
expression of $H$ explicitly depends only upon the locations and motion of
mass particles.

As with photons, the peculiar velocity of mass particles also cannot be used
in the definition of $H$. The peculiar velocity of a particle is defined as
its velocity relative to the local Hubble drift in which it is located. It
therefore depends on the Hubble drift and this requires, of course, that $H$
be defined before-hand. If the definition of $H$ is to be independent of
abstract concepts and depend only on observable physical quantities that
exist in the universe, the sole assumption on which the cosmological
kinematics of the Hubble parameter can be based is as follows: \textit{An
equation defining the magnitude of }$H$\textit{\ and describing its
kinematic behavior must only be a function of the physical vector lengths
that define the positions of all existing mass particles in the universe,
and the velocities associated with these lengths.} The definition of the
Hubble parameter $H$ will be directly determined here by explicit quantities
expressing the vector position $\mathbf{S}_{k}$ of each particle $k$, rather
than a general vector function $\mathbf{S}$\ that refers to a position in
space whether or not it defines the position of a particle. The lengths used
will be physically measurable lengths, which are the integrals of the space
components of the geodesic metric length $ds$ of general relativity existing
within what will be assumed is a universe generally consisting of flat
space. The physical lengths are generally not coordinate lengths and are
independent of abstract concepts used in coordinate systems. Curvature of
space or mass distributions in a flat universe can distort local space-time
and the lengths they contain; these distorted lengths are defined as
physical lengths as explained, for example, by Sokolnikoff [2]. Local
gravitational distortions of length are extremely small compared to the vast
cosmological distances in the universe so that they differ very slightly
from coordinate lengths, and physical lengths are essentially used here for
their conceptual importance.

\bigskip

\textbf{II. Kinematic and Gravitational\ Definitions of }$H$\textbf{\ }

\bigskip

A general definition of the Hubble parameter can now be found by assuming it
is a function of only the distances and velocities of all mass particles
that at the time of the origin of the universe are relatively few in number,
have all been created within a small finite volume of space by the radiation
field which may be, but is not necessarily, finite and which homogeneously
fills the universe. Only after the universe's initial expansion has occurred
would a large and continuous creation of particles take place throughout the
radiation universe and consequently produce a homogeneous distribution of
matter. Photon particles are not included in the definition of $H$ for the
reason explained in the introduction. A complete kinematic description of
the Hubble parameter was described by the author in a previous paper [3].
Because the cosmological distances between mass particles are large compared
their size, the universe will be considered to macroscopically consist of a
number $N$ of point-like mass particles of equal rest mass $m$. When $N$ is
large, any two unequal macroscopic masses could simply be considered to be
composed of different numbers of point-like masses. The rest mass and not
the relativistic mass is used because relativistic mass is a function of its
velocity relative to inertial systems which are to be defined here. The
vector positions $\mathbf{S}_{k}$ as used here will always represent the
distance of a particle $k$ having a rest mass $m$ within a spatially flat
universe which may contain intermittent local volumes of curved space-time
due to local gravitational distortions generated by stars or galaxies. It
represents the spatial three-dimensional metric length along a geodesic path
that is the physical length running from the center of the expansion, which
is the center of the initial mass distribution, to the position of particle $%
k$. Assuming the expansion initially begins within a very small volume of
space containing all the mass particles within an otherwise radiation
universe, a continuing process of particle creation throughout the radiation
universe would subsequently occur and can be shown to quickly reduce the
physical effects of the initial particles located at the center of the
expansion. According to the superposition of the two kinds of motion that
occur on the cosmological level, the velocity of a particle is then defined
as%
\begin{equation}
\mathbf{\dot{S}}_{k}=H\mathbf{S}_{k}+\mathbf{\dot{s}}_{k}.  \tag{1}
\end{equation}

\noindent Here, $\mathbf{\dot{s}}_{k}$ is the peculiar velocity of particle $%
k$ which would be observed as the particle's velocity relative to the local
inertial system that is itself at rest relative to the radial Hubble
velocity $H\mathbf{S}_{k}$ at the particle's location. The unit vector along
the direction of $\mathbf{S}_{k}$ is defined as $\mathbf{k}=\mathbf{S}%
_{k}/S_{k}$. The same letter $k$ is used here to represent both the unit
vector $\mathbf{k}$ as well as the subscript in the letter defining the
particle's position, however they are easily distinguished because one is a
vector and the other a scalar.

The time $t$ contained in the velocities is here defined as the cosmic time
measured by a clock that is a rest relative to the Hubble drift in which the
mass particle is located.\ As assumed in relativity theory and the
cosmological principle, the laws of physics are the same everywhere. It must
then be expected that time $t$ measured by a clock at rest in a galaxy would
flow at the same rate in every galaxy that is at rest relative to its local
Hubble drift. The Lorentz transformation and relativity then describe time
transformations among coordinate systems within a single galaxy, but not
between two galaxies having relative velocities generated by the Hubble
expansion and containing identical clocks. It is for these reasons that the
kinematic equations developed here are different from, and cannot have, the
covariant form of relativity theory.

As stated in the introduction, an expression for $H$ can be obtained which
is a function of only $\mathbf{S}_{k}$ and its derivative with respect to
time $\mathbf{\dot{S}}_{k}$. By taking the scalar product of both sides of
Eq. (1) with $\mathbf{k}$ and summing over all values of $k$ from $1$ to $N$%
, an expression for the Hubble parameter in the form 
\begin{equation}
H=(\Sigma \mathbf{\dot{S}}_{k}\cdot \mathbf{k})/(\Sigma \mathbf{S}_{k}\cdot 
\mathbf{k})  \tag{2a}
\end{equation}

\noindent is found when the important condition 
\begin{equation}
\Sigma \mathbf{\dot{s}}_{k}\cdot \mathbf{k=\,}0  \tag{2b}
\end{equation}

\noindent is satisfied by all the radial components of the peculiar
velocities $\mathbf{\dot{s}}_{k}$. The terms in Eqs. (2a) and (2b) are
scalar functions which represent only the radial components of $\mathbf{S}%
_{k}$, $\mathbf{\dot{S}}_{k}$, and $\mathbf{\dot{s}}_{k}$ due to the scalar
product with $\mathbf{k}$. Once the magnitude of $H$ is defined by Eq. (2a),
the magnitudes of the peculiar velocities $\mathbf{\dot{s}}_{k}$ are also
determined by Eq. (1) in terms of $\mathbf{\dot{S}}_{k}$ and $H\mathbf{S}%
_{k} $. The definition of $H$ in Eq. (2a) cannot contain the peculiar
velocities $\mathbf{\dot{s}}_{k}$ for the same reason that it cannot contain
references to photon velocities because both are defined relative their
inertial system's receding Hubble motion. The definition of the expansion
parameter $a(t)$ is given by $\mathbf{S}_{k}=\mathbf{h}_{k}a$, where $%
\mathbf{h}_{k}$ is a vector length that is independent\ of $a$ and is
associated with particle $k$ as $\mathbf{S}_{k}$ expands as a linear
function of $a$. When it is substituted into Eq. (2a) it reduces to the
general definition of the Hubble parameter $H=\dot{a}/a$ by recognizing that
the peculiar velocity is $\mathbf{\dot{s}}_{k}$ $=\mathbf{\dot{h}}_{k}a$
which in turn must satisfy Eq. (2b). Consequently Eqs. (2a) and (2b) are
then justified as a physically correct definition of $H$. Of course, if $%
\mathbf{\dot{S}}_{p}=H\mathbf{S}_{p}$ and $\mathbf{\dot{S}}_{q}=H\mathbf{S}%
_{q}$ are the Hubble velocities for particles $p$ and $q$ receding relative
to the center of the expansion, then the difference between these two
equations $\mathbf{\dot{S}}_{p}-\mathbf{\dot{S}}_{q}=H(\mathbf{S}_{p}-%
\mathbf{S}_{q})$ shows that the Hubble recession also occurs between any two
particles that have no peculiar velocities, and does not occur only relative
to the center of the expansion.

\ If particles are joined together to form particles of different masses,
these equations then contain weighted positions and velocities. When the
numerator and denominator in Eq. (2a) are each divided by $N$, it is seen
that the equation represents the average radial speed of the mass particles
divided by their average distance from the center of the expansion. When Eq.
(2b) is multiplied by the mass $m$ it has a form somewhat similar to the
conservation of momentum arising from Newton's third law of motion
concerning action and reaction. However, it is not a Newtonian equation
because each particle located at $\mathbf{S}_{k}$ has a different direction
for its unit vector $\mathbf{k}$. It then requires that the sum of only
radial momentums of the peculiar velocities vanish. This means that the
local radial interaction between particles cannot change the total radial
momentum of peculiar velocities or affect the magnitude of $H$.

The magnitude of $H$ is shown in [3] to be kinematically changed only by the
peculiar velocities $\mathbf{\dot{s}}_{k}$. Of course $H$ is also affected
by gravitation on a cosmological scale as described by general relativity.
Newtonian gravitational mechanics can yield the same exact cosmological
equations as found in general relativity, as was originally pointed out by
Milne [4] and\ Milne and McCrea [5]. The gravitational equations for $H$ are
found to have the general form

\begin{equation}
H^{2}=\frac{8\pi G}{3}\rho -\frac{c^{2}}{R_{0}^{2}a^{2}}+\frac{\Lambda }{3},
\tag{3}
\end{equation}

\noindent where $\rho $ is the density of mass and the equivalent mass of
radiation energy, $R_{0}$ is a scaling constant within the curvature term, $%
\Lambda $ is Einstein's cosmological constant, and $G$ is the gravitational
constant. By taking the time derivative of Eq. (3) and using Lema\^{\i}tre's
equation%
\begin{equation}
\dot{\rho}=-3H(\rho +p/c^{2}),  \tag{4}
\end{equation}

\noindent where $p$ is radiation pressure, the gravitational change in the
Hubble parameter is seen to be

\begin{equation}
\dot{H}=-4\pi G\left( \rho +\frac{p}{c^{2}}\right) +\frac{c^{2}}{%
R_{0}^{2}a^{2}}.  \tag{5}
\end{equation}

Equations (2a) and (2b) define the Hubble parameter in kinematic terms
involving only particle positions and velocities. Solutions to Eq. (5), with
the help of Eq. (4), also define the Hubble parameter in terms of
gravitation. Because the Hubble parameter is defined as $H=\dot{a}/a$ in
both cases, these two expressions for $H$ must have identical magnitudes.
This physical identity will be used to show that $\Lambda =0$ is a required
condition in the standard Friedman-Lema\^{\i}tre solution of the
cosmological equations in general relativity.

\bigskip

\noindent \textbf{III. PHYSICAL CONDITIONS GENERATING A BALANCED EXPANSION
WITH }$\mathbf{\Lambda =0}$

\bigskip

The Guth inflationary model involves a universe consisting of massless
particles and contains an initial inflationary expansion followed by a
second expansion of the kind that occurs in the Friedman-Lema\^{\i}tre
solution. This could occur when the mass particles generating the
inflationary expansion have a very short life-time so that it ends as a pure
radiation model containing no mass particles. At that instant the universe\
could transition into one with a set of newly created mass particles that
are different from the inflationary particles that had been created in a
much more dense universe than those which were created after the end of the
inflationary period, all of which are described here in terms of primed
letters. For simplicity the particles will be assumed in this case to form a
single small spherical shell of radius $S_{k}=r_{0}^{\prime }$ on which they
may or may not be uniformly distributed at the time of their creation.
However, each particle is created with an outwards moving initial radial
velocity $v_{0}^{\prime }$, with no tangential velocities. When all the
existing mass particles in the universe have the same positive radial
velocity $\dot{S}_{k}=v_{0}^{\prime }$, they would then not have peculiar
velocities because Eq. (2a)\ then defines $H=v_{0}^{\prime }/r_{0}^{\prime }$
where $\mathbf{\dot{s}}_{k}=0$ for all particles, which satisfies Eq. (2b).
In the absence of peculiar velocities no radiation pressure caused by
Doppler effects can act on the particles. When the energy density $u=\rho
c^{2}$ and the radiation pressure $p=u/3$ are substituted into Eq. (4) it is
seen that the equation is satisfied when the equivalent mass density of the
homogeneous and isotropic radiation is $\rho =\rho _{0}(a_{0}/a)^{4}$, where 
$\rho _{0}$ is the density when $a=a_{0}$.

When these expressions for $\rho $ and $u$ are substituted into Eq.(5) for
the case of a flat universe with $R_{0}=\infty $, and $a^{\prime }$ is used
to define the post-inflationary expansion parameter, the result is 
\begin{equation}
\dot{H}=-\kappa _{3}^{\prime }(a_{0}^{\prime }/a^{\prime })^{4},  \tag{6}
\end{equation}%
where $\kappa _{3}^{\prime }=(16/3)\pi G\rho _{0}^{\prime }$ with a
numerical value $\kappa _{3}^{\prime }=1.12\times 10^{54}s^{-2}$ when it is
assumed to contain the post-inflation equivalent mass density of $\rho
_{0}^{\prime }=10^{60}g/cm^{3}$. A solution to this equation occurring in
the standard model of relativity, where $H=\dot{a}/a$, is%
\begin{equation}
a^{\prime }=(\sqrt{2\kappa _{3}^{\prime }}\,t^{\prime
}+1)^{1/2}a_{0}^{\prime },  \tag{7}
\end{equation}

\noindent with $t^{\prime }=0$ occurring at the instant that the expansion
of the standard model began. At the time that this second expansion begins
the expansion parameter is $a_{0}^{\prime }=10^{30}a_{0}$, where $a=a_{0}$
at the time when inflation began.

At the end of the inflationary expansion, when the second (relativistic)
expansion begins, the time derivative of Eq. (7), defines a new value for $%
\dot{a}$ which is much smaller than its previous value that occurred at the
end of the inflationary expansion. Substituting Eq. (7) into the right-hand
side of Eq. (6) yields an equation which has the solution%
\begin{equation}
H=\sqrt{\kappa _{3}^{\prime }/2}\left( \sqrt{2\kappa _{3}^{\prime }}%
\,t^{\prime }+1\right) ^{-1}+\sqrt{\frac{\Lambda }{3}}  \tag{8}
\end{equation}

\noindent where $\Lambda $ is a cosmological constant. Equation (8)
describes $H$ as a function of time that originated from the use of Eq. (5),
while Eq. (2a) describes it as a function of the positions and velocities of
the mass particles. Because they both define the Hubble parameter as $H=\dot{%
a}/a$ both of these equations can be set equal to each other so that%
\begin{equation}
\frac{\Sigma \mathbf{\dot{S}}_{k}^{\prime }\cdot \mathbf{k}}{\Sigma \mathbf{S%
}_{k}^{\prime }\cdot \mathbf{k}}=\sqrt{\kappa _{3}^{\prime }/2}\left( \sqrt{%
2\kappa _{3}^{\prime }}\,t^{\prime }+1\right) ^{-1}+\sqrt{\frac{\Lambda }{3}}%
.  \tag{9}
\end{equation}%
At the time $t^{\prime }=0$ the identical radii and radial velocities $%
S_{k}=r_{0}^{\prime }$ and $\dot{S}_{k}=v_{0}^{\prime }$ are used for all
particles on the spherical shell. The parameter $H$ on the left-hand side of
Eq. (9) then becomes simply $v_{0}^{\prime }/r_{0}^{\prime }$. When $\Lambda
=0$ while substituting $\ \kappa _{3}^{\prime }=(16/3)\pi G\rho _{0}^{\prime
}$ into Eq. (9), it takes the form

\begin{equation}
\frac{v_{0}^{\prime }}{r_{0}^{\prime }}=\sqrt{\frac{\kappa _{3}^{\prime }}{2}%
}=\sqrt{\frac{8}{3}\pi G\rho _{0}^{\prime }}.  \tag{10}
\end{equation}%
\bigskip

The expression on the left-hand side of Eq. (9) is independent of the number
of mass particles $N$ that form the sphere \ Thus, even if only two
particles existed so that $N=2$, the form and behavior of Eq. (9) would be
unaltered.

It is important to recognize here that, as stated in the introduction, it is
only in the inertial systems that are at rest relative to their local Hubble
drift that the cosmological background radiation is observed to be
isotropic. The mass particles in Eq. (2a) determine the magnitude of $H$ and
therefore define the motion of these inertial systems and of course other
local inertial inertial can have constant velocities relative to them.
However, the conservation of energy occurs only within an inertial system.
When the first particles are created within a radiation universe, they also
define the motion of inertial systems, but only after $v_{0}^{\prime }$ and $%
r_{0}^{\prime }$are created. Without an existing inertial system, relative
to which mass particle velocities can be defined, the kinetic energy also
cannot be defined. Consequently the only energy that could be used by the
radiation to create mass particles would be the energy it used to create the
rest energy of the masses $E=mc^{2}$. Without a defined inertial system the
radiation could not possibly create any energy in addition to rest mass
energy which is independent of velocity and the inertial system in which it
is defined.

Now consider the physical meaning of Eq.(10). By squaring both sides of Eq.
(10) and then multiplying both sides by $mr_{0}^{^{\prime }2}/2$, while
recognizing that $M=(4/3)\rho _{0}^{\prime }\pi r_{0}^{\prime 3}$ is the
total equivalent mass of the radiation inside the spherical shell, it takes
the form 
\begin{equation}
\frac{1}{2}mv_{0}^{\prime 2}-\frac{GMm}{r_{0}^{\prime }}=0.  \tag{11}
\end{equation}

Eq. (11) states the that the total energy of a mass particle due to its
kinetic and gravitational potential energies is zero so that all of a mass
particle's energy is contained in its rest mass as $mc^{2}$. This is
obviously the expression for energy that occurs for a particle that has the
minimum velocity required for it to escape from a gravitational mass $M$ so
that $v_{0}^{\prime }\rightarrow 0$ as $r_{0}^{\prime }\rightarrow \infty $.
This would hold true for every particle in the particle shell. In this case
of mass particles expanding away from the center of the shell which contains
the isotropic radiation of density $\rho _{0}^{\prime }$, the Hubble
parameter $H\rightarrow 0$ as $r_{0}^{\prime }\rightarrow \infty $. As a
result the Hubble expansion is then balanced as would be expected when $%
\Lambda =0$.

The initial radius $r_{0}^{\prime }$ of the shell in Eq. (9) at the time $%
t^{\prime }=0$, when the relativistic expansion began at the end of the
inflationary expansion, depends on the initial velocity of the particles.
Assuming $v_{0}^{\prime }=c/10$ while $\kappa _{3}^{\prime }=1.12\times
10^{54}s^{-2}$,\ when the initial density $\rho _{0}^{\prime
}=10^{60}g/cm^{3}$in Eq. (10), yields $r_{0}^{\prime }=4.00\times 10^{-18}cm$
as the required radius that would create this balanced universe. If in
addition to the radial velocities, small tangential velocities had occurred,
it can be shown that such velocities could generate a kinematic effect that
would yield small values for $\Lambda $. The kinematics of both radial and
tangential velocities are discussed in a previous paper in [3].

\bigskip

\textbf{IV. Summary}\bigskip

The kinematic definition of the Hubble parameter in Eq.(2a) was shown to be
able to take the form $H=\dot{a}/a$ which is then physically identical to
the relativistic definition of the Hubble parameter. Equation (2a) defines
the Hubble parameter in terms of the positions and velocities of all mass
particles in the universe. The gravitational expression for $H$ found in the
in the standard model of relativity was set equal to the kinematic
expression for $H$ as done in Eq. (9) , which occurs after the inflationary
expansion. The kinematic definition of the Hubble parameter defines the rate
of the universe's expansion, and energy conservation occurs only within an
inertial coordinate system. The isotropic form of the cosmic background
radiation is observed only within an inertial system that is at rest
relative to its local Hubble drift so that $H$ kinematically defines
inertial systems at rest relative to the Hubble drift. When the expansion is
about to be generated by the first existing mass particles begins, these
particles are forming in a universe with no defined inertial system or
systems so that velocity, conservation of energy, or even energy itself
cannot be defined relative to an inertial system. The universe's radiation
then only creates the rest mass energy of the particles which is independent
of both a particle's velocity and its kinetic energy. It requires in turn
that the total kinetic and gravitational energy of the particle be zero as
shown in Eqs.(10) and (11) where $\Lambda =0$. This condition remains in
effect while additional mass particles are subsequently created\ within
their respective local inertial systems that move with the Hubble motion
throughout the expanding universe.

\bigskip

\noindent \lbrack 1] A. H. Guth, Phys. Rev. D \textbf{23,} 347 (1981).

\noindent \lbrack 2] I. S. Sokolnikoff, \textit{Tensor Analysis Theory and
Applications to Geometry and Mechanics of Continua} (Wiley and Sons, \ \ \
1964), pp. 121-122.

\noindent \lbrack 3] D. Savickas, Int. J. Mod. Phys. D \textbf{2}, 197
(1993).

\noindent \lbrack 4] E.A. Milne, Quart.\ J. Math. \textbf{5}, 64 (1934).

\noindent \lbrack 5] W. H. McCrea and E. A. Milne, Quart. J. Math \textbf{5}%
, 73 (1934).

\end{document}